\documentclass[iop,apj,tighten]{emulateapj}

\usepackage{lscape}
\usepackage{color}

\newcommand{\lya}{Ly$\alpha$}
\newcommand{\lyb}{Ly$\beta$}

\newcommand{\hi}{\ion{H}{1}}

\newcommand{\oi}{\ion{O}{1}}
\newcommand{\ci}{\ion{C}{1}}
\newcommand{\cii}{\ion{C}{2}}

\newcommand{\civ}{\ion{C}{4}}
\newcommand{\siii}{\ion{Si}{2}}

\newcommand{\siiv}{\ion{Si}{4}}

\newcommand{\mgii}{\ion{Mg}{2}}
\newcommand{\feii}{\ion{Fe}{2}}
\newcommand{\mhi}{{\rm H \; \mbox{\tiny I}}}
\newcommand{\moi}{{\rm O \; \mbox{\tiny I}}}
\newcommand{\msiii}{{\rm Si \; \mbox{\tiny II}}}
\newcommand{\mfeii}{{\rm Fe \; \mbox{\tiny II}}}
\newcommand{\mmgii}{{\rm Mg \; \mbox{\tiny II}}}

\newcommand{\mcii}{{\rm C \; \mbox{\tiny II}}}

\newcommand{\kms}{km~s$^{-1}$}

\shorttitle{Fe and $\alpha$-elements at $z \sim 5$-6} 

\shortauthors{Becker et al.}

\begin{document}

\title{Iron and $\alpha$-element Production in the First One Billion Years after the Big Bang\altaffilmark{1,2,3}}

\author{George D. Becker\altaffilmark{4}, Wallace L. W. Sargent\altaffilmark{5}, Michael Rauch\altaffilmark{6}, Robert F. Carswell\altaffilmark{4}}

\altaffiltext{1}{The observations were made in part at the W.M. Keck Observatory, which is operated as a scientific partnership between the California Institute of Technology and the University of California; it was made possible by the generous support of the W.M. Keck Foundation.}
\altaffiltext{2}{This paper includes data gathered with the 6.5 meter Magellan Telescopes located at Las Campanas Observatory, Chile.}
\altaffiltext{3}{Based in part on observations made with the Very Large Telescope, operated by the European Southern Observatory at Paranal Observatory, Chile, under proposal ID 084.A-0574.}
\altaffiltext{4}{Kavli Institute for Cosmology and Institute
  of Astronomy, Madingley Road, Cambridge, CB3 0HA, UK;
  gdb@ast.cam.ac.uk, acalver@ast.cam.ac.uk}
\altaffiltext{5}{Palomar Observatory, California Institute of
  Technology, Pasadena, CA 91125, USA; wws@astro.caltech.edu}
\altaffiltext{6}{Carnegie Observatories, 813 Santa Barbara Street,
  Pasadena, CA 91101, USA; mr@obs.carnegiescience.edu}

\slugcomment{ApJ, Accepted November 19, 2011}

\begin{abstract}

We present measurements of carbon, oxygen, silicon, and iron in quasar absorption systems existing when the universe was roughly one billion years old.  We measure column densities in nine low-ionization systems at $4.7 < z < 6.3$ using Keck, Magellan, and VLT optical and near-infrared spectra with moderate to high resolution.  The column density ratios among \cii, \oi, \siii, and \feii\  are nearly identical to sub-DLAs and metal-poor ($[{\rm M/H}] \le -1$) DLAs at lower redshifts, with no significant evolution over $2 \lesssim z \lesssim 6$.  The estimated intrinsic scatter in the ratio of any two elements is also small, with a typical r.m.s. deviation of $\lesssim 0.1$ dex.  These facts suggest that dust depletion and ionization effects are minimal in our $z > 4.7$ systems, as in the lower-redshift DLAs, and that the column density ratios are close to the intrinsic relative element abundances.  The abundances in our $z > 4.7$ systems are therefore likely to represent the typical integrated yields from stellar populations within the first gigayear  of cosmic history.  Due to the time limit imposed by the age of the universe at these redshifts, our measurements thus place direct constraints on the metal production of massive stars, including iron yields of prompt supernovae.  The lack of redshift evolution further suggests that the metal inventories of most metal-poor absorption systems at $z \gtrsim 2$ are also dominated by massive stars, with minimal contributions from delayed Type Ia supernovae or AGB winds.  The relative abundances in our systems broadly agree with those in very metal-poor, non-carbon-enhanced Galactic halo stars.  This is consistent with the picture in which present-day metal-poor stars were potentially formed as early as one billion years after the Big Bang.

\end{abstract}


\section{Introduction}\label{sec:intro}

Metal abundances provide some of our most valuable insights into star formation and stellar evolution.  The elements produced by a single star result from a variety of nuclear reactions occurring over its life and death, the details of which will depend on the star's mass and initial chemical composition, and whether it is a member of a binary or higher multiple system.  Measurements of element abundance ratios can therefore be used to place constraints on stellar evolution models.  In turn, understanding stellar evolution and nucleosynthetic yields can be used to infer the star formation history and mass assembly of galaxies such as the Milky Way \citep[for reviews, see][]{mcwilliam1997,matteucci2008}.

Of particular interest are the yields from the first few generations of stars.  The integrated metal production of the so-called ``Pop III'' stars, which formed out of pristine material, will depend strongly on their initial mass function (IMF).  In the conventional picture, where Pop III stars are predominantly massive due to a lack of metal-line cooling in the collapsing gas clouds \citep[e.g.,][]{bromm2004}, their nucelosynthetic signature may be highly distinct from later Pop II stars \citep[e.g.,][]{heger2002,nomoto2006}.  Recent simulations, however, suggest that fragmentation due to turbulence may produce a Pop III IMF that is less top heavy than  previously thought \citep{clark2011}.  In either case, the metal abundances in low-mass Pop II stars that survive from this period should encode information on the nature of the first stars.  In addition, early abundance patterns should reflect some of the key, and as yet uncertain, characteristics governing star formation in the early universe, including the IMF.

A classic approach to studying early star formation is to measure metal abundances in very metal-poor stars.  Significant numbers of objects with metallicities as low as ${\rm [Fe/H]} \sim -5.5$ \citep{christlieb2002,frebel2005} have now been identified in the Milky Way and its satellite galaxies.  Indeed, metal-poor stars have produced a diverse array of insights, from the nucleosynthetic yields of low metallicity stars to the mass assembly of the Milky Way \citep[for reviews, see][]{beers2005,frebel2010}.  

Using metal-poor stars to determine early abundance patterns involves a number of challenges, however.  First, while highly precise differential stellar abundances can be determined under the simplifying assumptions of local thermodynamic equilibrium (LTE) and one-dimensional model atmospheres, measuring {\it absolute} abundances generally requires that non-LTE and 3D effects be considered.  These can introduce large and sometimes uncertain corrections, depending on the element and the choice of abundance indicator \citep[for a review, see][]{asplund2005a}.  Mixing of material within the stellar interior, moreover, may contaminate photospheric abundances such that they do not reflect the composition of the gas out of which the star was formed.  Second, although metal-poor stars are presumed to be old, the fact that these stars tend to be found in the field, rather than in clusters, means that age estimates cannot be derived from single stellar population modeling.  In some cases, radioactive dating has confirmed the old ages for metal-poor stars with enhanced $r$-process elements \citep{hill2002,sneden2003,frebel2007a}, but this has so far been done for only a handful of stars.  Uncertainties in the ages of most individual metal-poor stars therefore remain considerable, although isochrome-based methods suggest a typical age of $\sim$10-12 Gyr for halo stars with $-3.0 < [{\rm Fe/H}] < -1.3$ \citep{jofre2011}.  Finally, a number of metal-poor stars, including a large fraction of those with ${\rm [Fe/H]} < -3$, display ``anomalous'' abundance features such as strong enhancements in carbon, oxygen, and nitrogen \citep{beers1992,beers2005,cohen2005,frebel2005,frebel2006,norris2007}.  The enhanced fraction increases towards lower metallicities, with three out of the four known stars with ${\rm [Fe/H]} < -4.5$ being carbon-enhanced \citep[an exception having only recently been identified by][]{caffau2011}.  The origin of this enhancement is currently debated \citep[e.g.,][]{beers2005,iwamoto2005,meynet2006,carollo2011}.  If it results from mass transfer from a binary companion, however \citep{suda2004,campbell2010}, or if the gas needs to reach a threshold level of carbon and/or oxygen before it can cool and fragment \citep{bromm2003,frebel2007b}, then the abundances in these stars may not reflect  typical ISM abundances at low metallicities.

Quasar metal absorption lines provide an attractive complement to metal-poor stars for measuring primitive abundances, for several reasons.  Quasar absorption lines should integrate over larger spatial areas than the birthplaces of individual stars, and so may more accurately reflect mean ISM abundances.  A potentially more significant advantage, however, is that the conversion from observed line properties to total element abundances is often straightforward in quasar absorption systems.  This is particularly true for the high column density absorbers known as damped \lya\ systems (DLAs, $\log{N({\mbox \hi})} \ge 20.3$), which probe the interstellar media of intervening galaxies along quasar lines of sight \citep[e.g.,][]{wolfe2005}.  These systems are highly optically thick to hydrogen-ionizing photons, which ensures that nearly all of the elements in the ``neutral''-phase gas are in the lowest ionization state that has an ionization potential $\ge 1$~Ryd (e.g., carbon, silicon, magnesium, and iron will be singly ionized, while oxygen will be fully neutral).  Furthermore, the densities are low enough that nearly all of the ions will be in the ground state.  The net result is that, in the simplest case, abundances for a given element can be computed directly from the column density measurements of ions in a single electronic state.\footnote{For this reason we adopt the common convention among quasar absorption line studies of referring to an ion by its absorption spectrum.  Hence, $N({\rm C~II})$ is the column density of carbon ions producing C~II absorption, or equivalently, the column density of C$^{+1}$ ions.}  Complications do arise at lower \hi\ column densities, where ionization effects start to become significant, or if there is significant depletion onto dust grains.  Lines may also become saturated, preventing accurate column density measurements.  We will return to these points later in the paper.  

An equally important advantage of quasar absorption lines is the direct access they provide to abundances in the early universe.  With the discovery of metal-enriched absorbers out to redshift six \citep{becker2006,becker2009,becker2011b,rw2006,rw2009,simcoe2006b,simcoe2011a}, it is now possible to measure abundances in gas within one billion years of the Big Bang.  At this point, the finite age of the universe requires that the metals we observe must have been produced only by massive stars that exploded as core-collapse (Type II and Ib/c) and/or prompt Type Ia supernovae.\footnote{Here we distinguish between a potential ``prompt'' contribution to the Type Ia rate from high-mass stars, and a ``delayed'' contribution from evolved low-mass stars \citep{dallaporta1973,oemler1979,mannucci2005,scannapieco2005,matteucci2006}.}  Indeed, although the universe is $\sim$1 Gyr old at $z \sim 5$-6, current estimates suggest that the cosmic stellar mass density at these redshifts doubles on timescales of $\sim$300 Myr \citep{gonzalez2011}.  It is likely, therefore, that the metals seen in absorption at $z \sim 5$-6 are often less than a few hundred million years old.  High-redshift quasar absorption lines thus provide unique constraints on nucleosynthesis in massive (and presumably metal-poor) stars.  At the same time, they provide insights into the types of stars that ended the cosmic Dark Ages.

In this paper we present measurements of the relative abundances of $\alpha$-elements (carbon, oxygen, and silicon) and iron in a sample of nine quasar absorption line systems spanning $4.7 < z < 6.3$.  These are the highest-redshift systems for which such measurements have been made, and the first within one billion years of the Big Bang.  They also represent nearly all of the currently known low-ionization absorbers at $z \sim 5$-6, and thus constitute an unbiased sample with regards to their metallicity.  In Section~\ref{sec:data} we describe the optical and near-infrared spectra used for this work.  In Section~\ref{sec:columns} we present the column density measurements for individual systems.  We then use these to compute relative abundances in Section~\ref{sec:abundances}, and compare the results to lower-redshift quasar absorption systems.  In Section~\ref{sec:discussion} we discuss the evidence that the relative abundances in these systems represent the typical integrated yields of low metallicity, massive stars in the early universe.  We also compare our measurements to the relative abundances in metal-poor Galactic halo stars.  We summarize our conclusions in Section~\ref{sec:conclusions}.  

Throughout the paper we express logarithmic relative abundances with respect to the solar values as $[{\rm X/Y}] = \log{(N_{\rm X}/N_{\rm Y})} - \log{(n_{\rm X}/n_{\rm Y})_{\odot}}$, where $N_{\rm X}$ is the column density of element X.  All quantities are computed with respect to the solar photospheric values from \citet{asplund2009}.

\section{The Data}\label{sec:data}

\begin{deluxetable}{lcccc}
   \tabletypesize{} 
   \tablewidth{3.0in}
   \centering
   \tablecolumns{4}
   \tablecaption{Data} 
   \tablehead{\colhead{QSO} & 
              \colhead{$z_{\rm em}$} &
              \colhead{Instrument} &
	      \colhead{$t_{\rm exp}$} \\
               & & & (hrs) } 
   \startdata
   SDSS~J0040$-$0915  &  4.98  &   MIKE       &  8.3 \\
                      &        &   NIRSPEC    & 11.0 \\
   SDSS~J0231$-$0728  &  5.42  &   X-Shooter  &  6.0 \\
   SDSS~J1208$+$0010  &  5.27  &   X-Shooter  &  9.0 \\
   SDSS~J0818$+$1722  &  6.00  &   NIRSPEC    & 25.5 \\
   SDSS~J1148$+$5251  &  6.42  &   NIRSPEC    &  7.5 \\
   \vspace{-0.1in}
   \enddata
   \tablecomments{Keck/HIRES optical spectra for SDSS~J0231$-$0728,
      SDSS~J0818$+$1722, and SDSS~J1148$+$5251 were presented in
      \citet{becker2006,becker2011b}.}
   \label{tab:obs}
\end{deluxetable}

The data used for this work include optical and near-infrared spectra of five QSOs at $4.98 \le z_{\rm em} \le 6.42$.  Metal absorption line measurements from Keck/HIRES (FWHM = 6.7~\kms) optical spectra of SDSS~J0231$-$0728, SDSS~J0818$+$1722, and SDSS~J1148$+$5251  were presented in \citet{becker2006,becker2011b}.  The remaining observations are summarized in Table~\ref{tab:obs}.  Optical data from Magellan/MIKE (FWHM = 13.6~\kms) for SDSS~J0040$-$0915 were also used in \citet{becker2011a}; however, the metal absorption lines are presented for the first time here.

We obtained near-infrared spectra of SDSS~J0040$-$0915, SDSS~J0818$+$1722, and SDSS~J1148$+$5251 with Keck/NIRSPEC \citep{mclean1998} between December 2009 and March 2010.  Each object was observed using a single setting in either the Nirspec-5 (1.41-1.81~$\mu$m; SDSS~J0040$-$0915 and SDSS~J0818$+$1722) or Nirspec-6 (1.56-2.32~$\mu$m; SDSS~J1148$+$5251) band.  The settings were chosen to cover at least one strong \feii\ line.  The data were reduced using a suite of custom routines, described in \citet{becker2009}.  These were designed to optimally subtract the sky background and extract the one-dimensional spectra, while minimizing the noise due to a variety of detector issues.  Corrections for telluric absorption were made using the atmospheric transmission spectrum of \citet{hinkle2003}.  The spectral resolution was measured from telluric absorption lines in standard star spectra, and was found to be $\sim$15-16~\kms\ (FWHM).

Optical and near-infrared spectra of SDSS~J0231$-$0728 and SDSS~J1208$+$0010 were obtained with VLT/X-Shooter \citep{sdodorico2006} between October 2009 and March 2010.  Slit widths of 0\farcs9 were used in both the VIS and NIR arms.  The data were reduced using a custom pipeline similar to that used for the NIRSPEC data, and atmospheric absorption corrections were again made using a model.  The resolution, measured from telluric lines in the quasar spectra, was found to be $\sim$23 ($\sim$32)~\kms\ in the VIS arm, and $\sim$31 ($\sim$34)~\kms\ in the NIR arm for SDSS~J0231$-$0728 (SDSS~J1208$+$0010).  Resolutions higher than the nominal values were achieved due to good seeing.

\section{Column Density Measurements}\label{sec:columns}

Column densities for \cii, \oi, and \siii\ were measured from optical spectra, while \feii\ measurements were mostly made from strong lines in the infrared.  The measurements were done using two techniques.  In most cases where the system appeared to have a single, narrow component, Voigt profiles were fit using {\sc vpfit}.  When fitting, we explicitly tied together the Doppler parameters for all ions, which assumes the lines are  turbulently broadened.  For the remaining systems we used the apparent optical (AOD) method \citep{savage1991}, which calculates a total column density by summing the optical depth over all pixels within a specified velocity interval.  For noisy data (e.g., the HIRES data for SDSS~J0231$-$0728), we first smoothed the spectrum by convolving the flux with a Gaussian kernel with FWHM equal to one half the instrumental resolution before measuring optical depths.  This has only a minor effect on the spectral resolution, but prevents individual pixels from dominating the optical depth by taking advantage of the fact that the absorption lines are expected to have a finite width.  For further discussion, see \citet{becker2011b}.  

Notes on individual systems are given below.  A summary of the measured column densities is given in Table~\ref{tab:columns}.

\subsection{SDSS~J0040$-$0915, $z=4.7393$}

\begin{figure*}
   \epsscale{0.8} 
   \centering 
   \plotone{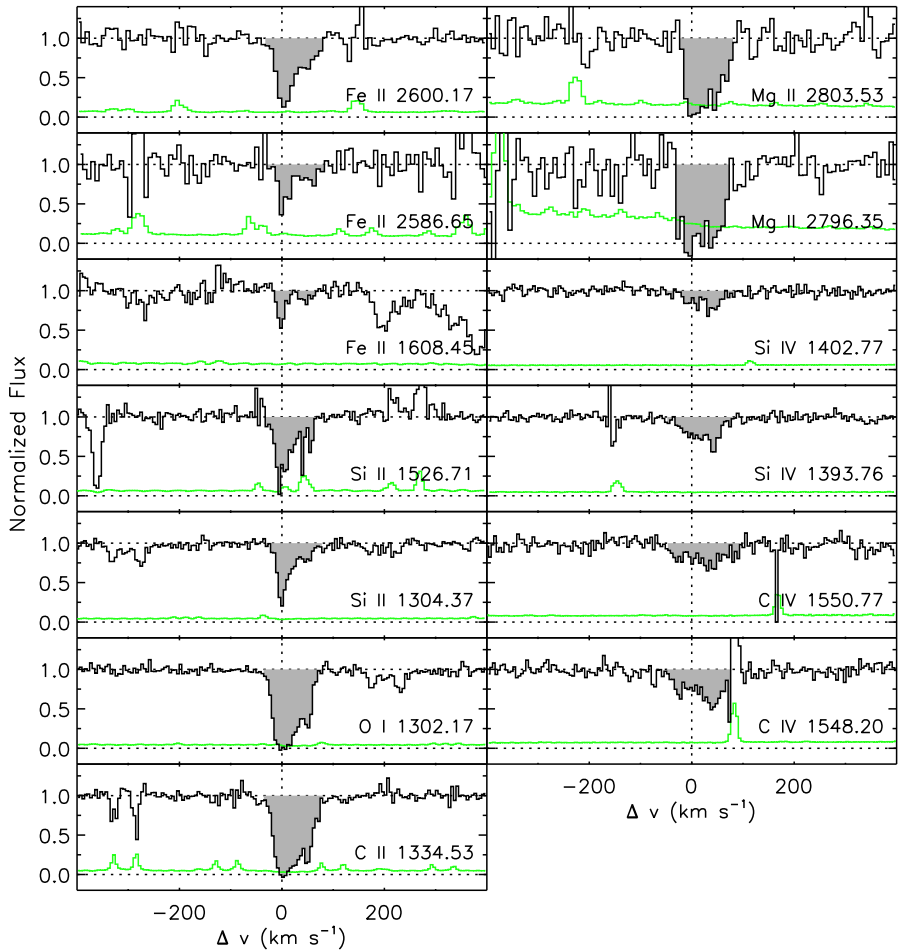}
   \caption{Metal lines for the $z= 4.7393$ system towards SDSS~J0040$-$0915.  Data covering transitions with rest wavelength up to 1608~\AA\ are from MIKE, while the remainder are from NIRSPEC.  The normalized flux for each ion is plotted at the nominal redshift of the system.  Shaded regions show detected transitions.  Peaks in the error array, which is plotted along the bottom of each panel, indicate regions affected by skylines, for example in the red wings of Si~II~1526 and C~IV~1548.
     \label{fig:0040_z4.7393}}
\end{figure*}

\begin{figure}
   \epsscale{0.93} 
   \centering 
   \plotone{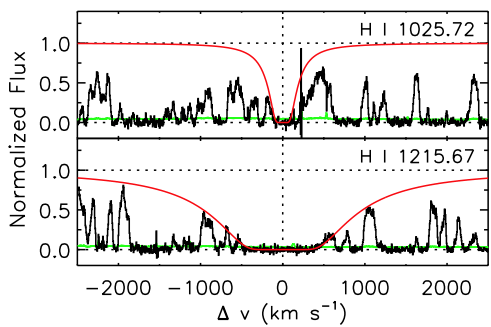}
   \caption{\lya\ and \lyb\ transitions for the $z= 4.7393$ system towards SDSS~J0040$-$0915.  The data are from MIKE.  Absorption profiles are not fits, but show the maximum allowed H~I column density, $\log{N_{\mhi}} = 20.4~({\rm cm^{-2}})$.
     \label{fig:0040_z4.7393_hi}}
\end{figure}

This system shows mildly saturated \cii~$\lambda$1334 and \oi$\lambda$1302, from which we can only derive lower limits on the column densities.  Multiple weak or moderate lines of \siii\ and \feii\ are present, however.  We also cover \mgii\ in the NIRSPEC data, although it is saturated.  Moderate \civ\ and \siiv\ absorption appears, but these lines are likely to arise from a separate, high-ionization phase \citep[e.g.,][]{fox2007a}.  The metal lines for this system are shown in Figure~\ref{fig:0040_z4.7393}.  At this redshift, the \lya\ forest is highly absorbed.  We can, however, set an upper limit on the \hi\ column density for this system from \lya\ and \lyb\ ($\log{N_{\mhi}} \le 20.4~({\rm cm^{-2}})$; Figure~\ref{fig:0040_z4.7393_hi}).

\subsection{SDSS~J1208$+$0010, $z=5.0817$}

\begin{figure*}
   \epsscale{0.8} 
   \centering 
   \plotone{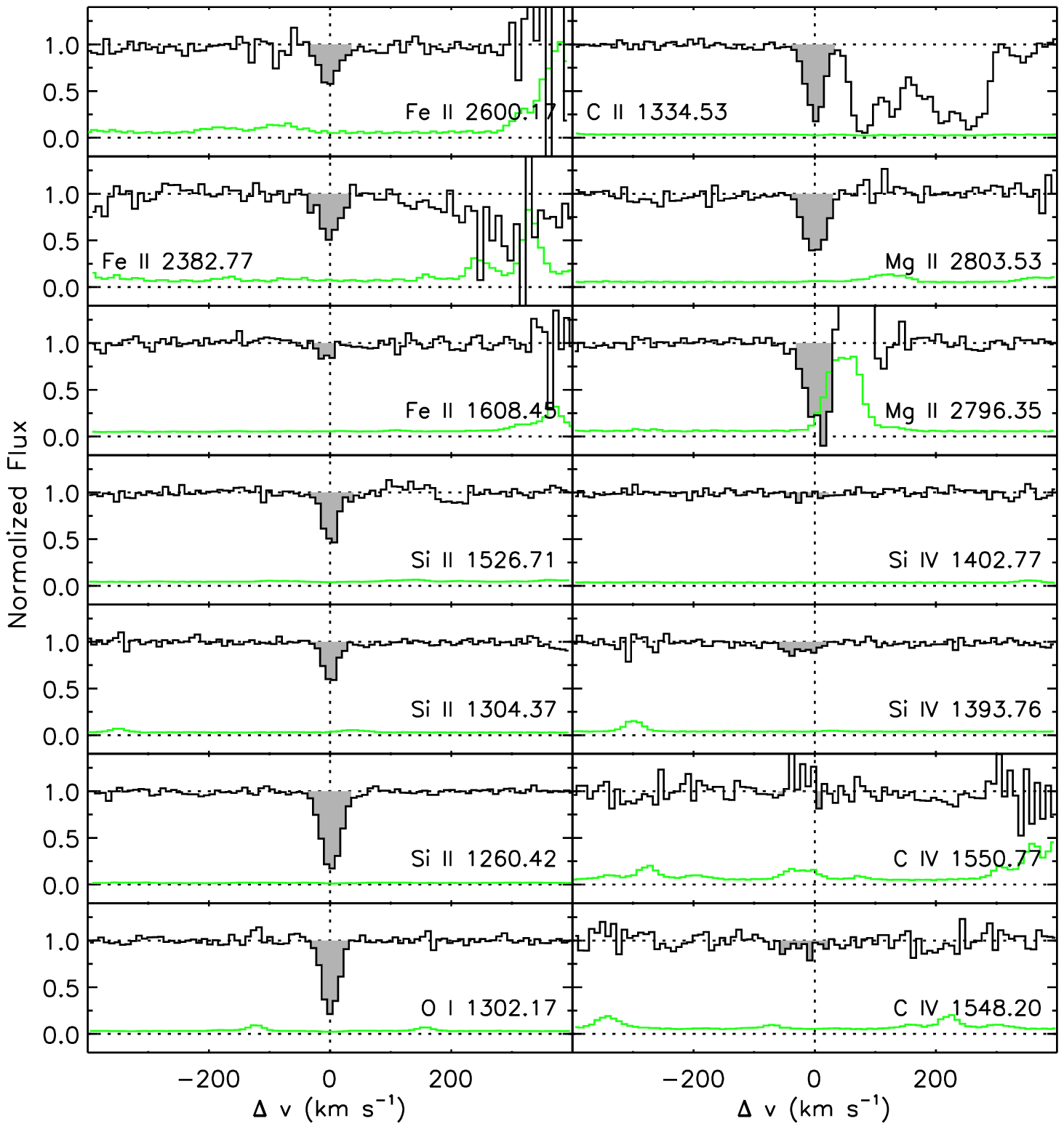}
   \caption{Metal lines for the $z= 5.0817$ system towards SDSS~J1208$+$0010.  All data are from X-Shooter.  Lines are as in Figure~\ref{fig:0040_z4.7393}.
     \label{fig:1208_z5.0817}}
\end{figure*}

\begin{figure}
   \epsscale{0.93} 
   \centering 
   \plotone{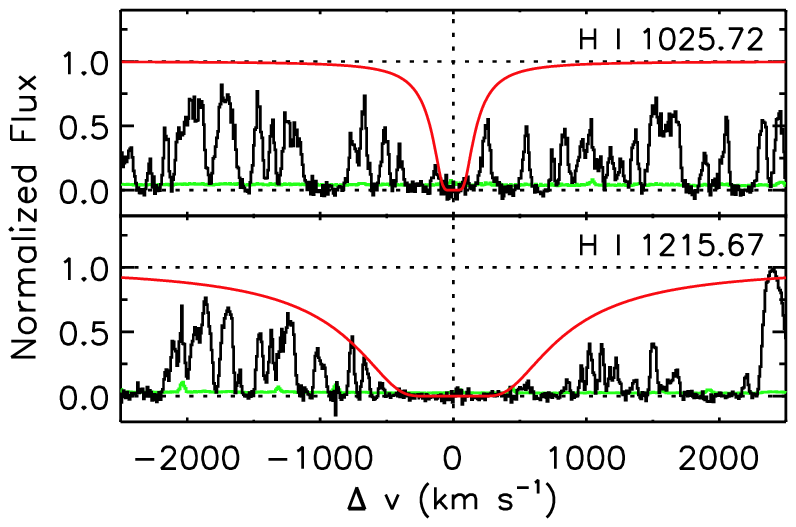}
   \caption{\lya\ and \lyb\ transitions for the $z= 5.0817$ system towards SDSS~J1208$+$0010.  The data are fromX-Shooter.  Absorption profiles are not fits, but show the maximum allowed H~I column density, $\log{N_{\mhi}} = 20.3~({\rm cm^{-2}})$.    Although the flux exceeds the absorption profile far out in the red wing of \lya, this discrepancy strongly depends on the uncertain continuum level.
     \label{fig:1208_z5.0817_hi}}
\end{figure}

This system appears to have a single component.  It is unresolved in the X-Shooter data, however, and we lack supplementary high-resolution spectra.  We have therefore attempted to obtain column densities by fitting Voigt profiles, with the $b$-parameter constrained by simultaneously fitting multiple \siii\ lines with different oscillator strengths.  A single-component fit gives $b = 9.5 \pm 0.2$~\kms, with which column densities for all ions can then be derived.  The column densities for \siii\ and \feii\ are likely to be robust, as we have at least one line in each case that are on the linear part of the curve of growth.  \cii\ and \oi\ are more problematic, however, as there could be hidden saturation issues.  We found that we could match the observed profiles of these ions with multiple, blended components that had a total column density significantly greater than that of the best-fit single-component fit.  Lower limits are therefore given for \cii\ and \oi, as well as for \mgii, for which the $\lambda$2796 line is strongly affected by sky-line residuals.  Weak high-ionization lines (\civ\ and \siiv) are detected for this system.  All of the metal lines are plotted in Figure~\ref{fig:1208_z5.0817}.  The regions of the forest covering \lya\ and \lyb\ are shown in Figure~\ref{fig:1208_z5.0817}.  The upper limit on the \hi\ column density is $\log{N_{\mhi}} \le 20.3~({\rm cm^{-2}})$.

\subsection{SDSS~J0231$-$0728, $z=5.3380$}

\begin{figure*}
   \epsscale{0.8} 
   \centering 
   \plotone{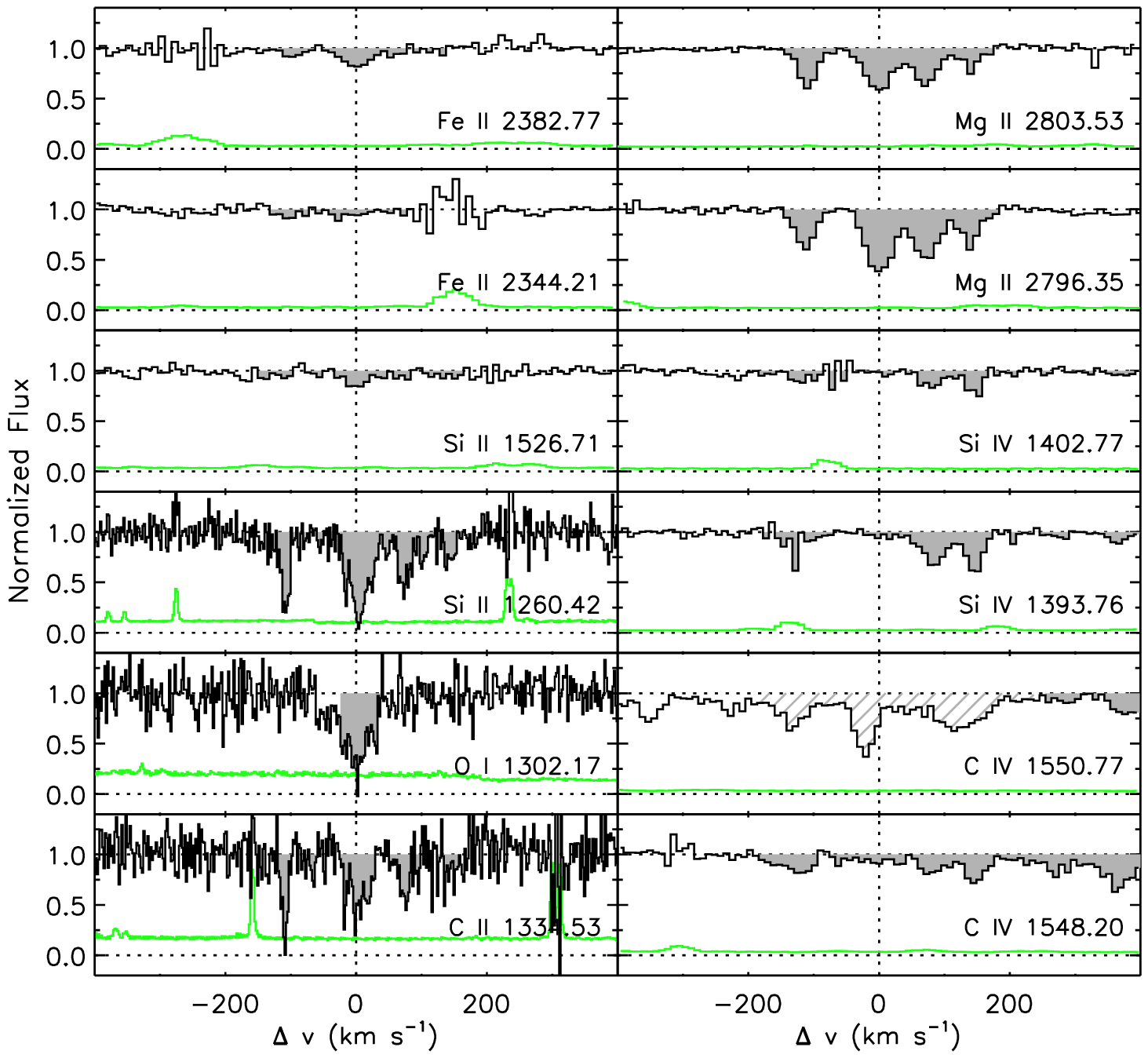}
   \caption{Metal lines for the $z=5.3380$ system towards SDSS~J0231$-0728$.  Data covering transitions with rest wavelength up to 1334~\AA\ are from HIRES, and were re-reduced from \citet{becker2006}.  The remaining data are from X-Shooter.  Lines are as in Figure~\ref{fig:0040_z4.7393}.  C~IV~1551 is blended with unrelated absorption lines.
     \label{fig:0231_z5.3380}}
\end{figure*}

The HIRES data for this system was originally presented in \citet{becker2006}.  We have re-reduced the data to take advantage of a number of improvements in the pipeline, and have re-measured column densities for \cii, \oi, and \siii.  Lines of \feii\ and \mgii, as well \siii~$\lambda$1526, are covered in the X-Shooter data.  \feii\ is detected in two lines with high oscillator strengths, $\lambda$2344 and $\lambda$2382.  We note that these lines are often badly saturated in lower-redshift DLAs and sub-DLAs, but are quite weak here.  Extended \civ\ and \siiv\ are present, but these are weak at the velocity of the strongest low-ionization component, where \oi\ is detected.  The metal lines are plotted in Figure~\ref{fig:0231_z5.3380}.  The upper limit on the \hi\ column density is $\log{N_{\mhi}} \le 20.3~({\rm cm^{-2}})$ \citep{becker2006}.

\subsection{SDSS~J0818$+$1722, $z=5.7911$}

\begin{figure}
   \epsscale{0.93} 
   \centering 
   \plotone{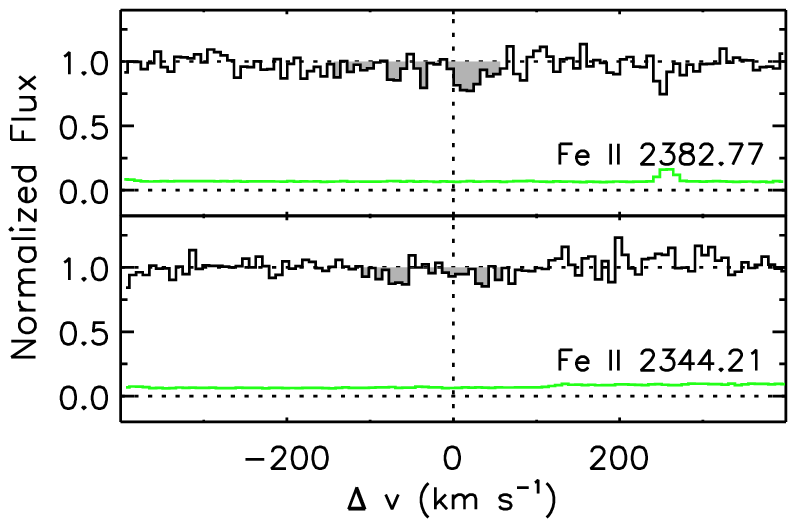}
   \caption{NIRSPEC data covering Fe~II for the $z=5.7911$ system towards SDSS~J0818$+$1722.  Lines are as in Figure~\ref{fig:0040_z4.7393}.  Data covering other ions for this system were presented in \citet{becker2011b}.
     \label{fig:0818_z5.7911}}
\end{figure}

The HIRES data for this system were presented in \citet{becker2011b}.  \feii\ is detected in the NIRSPEC data for two lines, $\lambda$2344 and $\lambda$2382.  Although the absorption is relatively weak, these show the same extended velocity structure present in the other low-ionization lines \citep{becker2011b}.  The \feii\ lines are shown in Figure~\ref{fig:0818_z5.7911}.

\subsection{SDSS~J0818$+$1722, $z=5.8765$}

\begin{figure}
   \epsscale{0.93} 
   \centering 
   \plotone{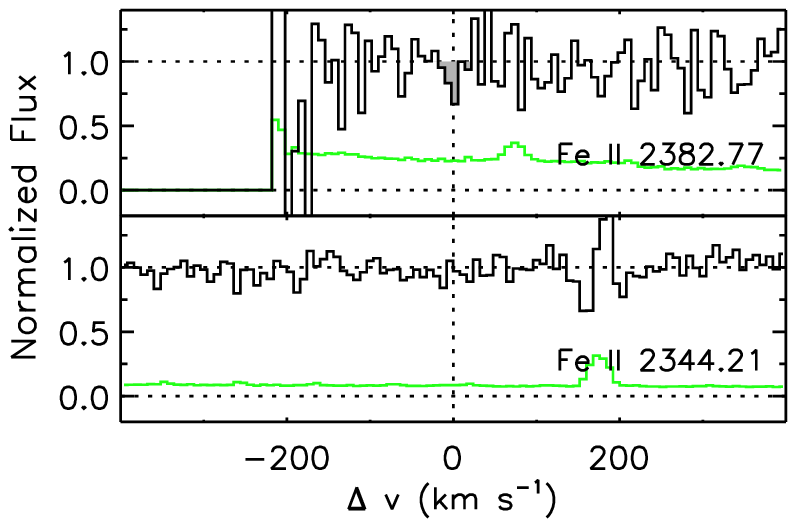}
   \caption{NIRSPEC data covering Fe~II for the $z=5.8765$ system towards SDSS~J0818$+$1722.  The absorption for Fe~II~$\lambda$2382 is not a significant detection.  Lines are as in Figure~\ref{fig:0040_z4.7393}.  Data covering other ions for this system were presented in \citet{becker2011b}.  
     \label{fig:0818_z5.8765}}
\end{figure}

The HIRES data for this system were presented in \citet{becker2011b}.  Although we cover \feii~$\lambda$2344 and $\lambda$2382 in the NIRSPEC data, no significant absorption is detected.  The expected strongest line, $\lambda$2382, unfortunately falls in a region of the spectrum with a low signal-to-noise ratio.  An upper limit is therefore given for $\log{N_{\mfeii}}$ in Table~\ref{tab:columns}.  The regions of the spectrum covering these lines are plotted in Figure~\ref{fig:0818_z5.8765}.

\subsection{SDSS~J1148$+$5251, $z=6.0115$}

\begin{figure}
   \epsscale{0.93} 
   \centering 
   \plotone{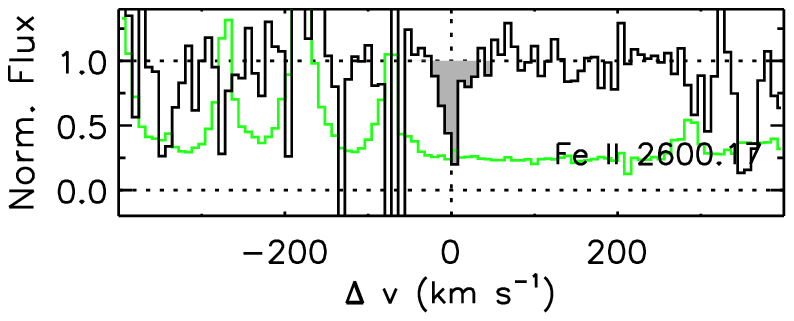}
   \caption{NIRSPEC data covering Fe~II for the $z=6.0115$ system towards SDSS~J1148$+$5251.  The Fe~II~2600 line falls redward of a region strongly affected by skylines, and is clearly detected.  Lines are as in Figure~\ref{fig:0040_z4.7393}.  Data covering other ions for this system were presented in \citet{becker2006,becker2011b}.    
     \label{fig:1148_z6.0115}}
\end{figure}

The HIRES data for this system were presented in \citet{becker2006} and re-analyzed in \citet{becker2011b}.  \feii~$\lambda$2600 is detected in the NIRSPEC data.  Although it falls adjacent to a region strongly contaminated by skylines, the \feii\ line is not affected, as shown in Figure~\ref{fig:1148_z6.0115}.

\subsection{SDSS~J1148$+$5251, $z=6.1312$}

\begin{figure}
   \epsscale{0.93} 
   \centering 
   \plotone{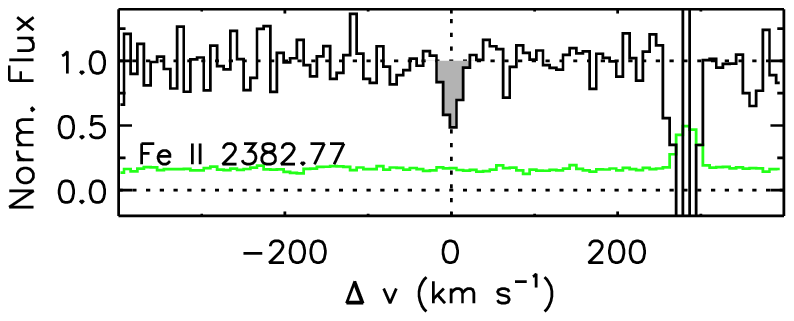}
   \caption{NIRSPEC data covering Fe~II for the $z=6.1312$ system towards SDSS~J1148$+$5251.  Lines are as in Figure~\ref{fig:0040_z4.7393}.  Data covering other ions for this system were presented in \citet{becker2006,becker2011b}.      
     \label{fig:1148_z6.1312}}
\end{figure}

The HIRES data for this system were presented in \citet{becker2006} and re-analyzed in \citet{becker2011b}.  We clearly detect \feii~$\lambda$2382 in the NIRSPEC data.  This line is plotted in Figure~\ref{fig:1148_z6.1312}.

\subsection{SDSS~J1148$+$5251, $z=6.1988$}

The HIRES data for this system were presented in \citet{becker2006} and re-analyzed in \citet{becker2011b}.  This is the weakest of our $z > 4.7$ systems, and although we cover \feii~$\lambda$2344 in the NIRSPEC data, the spectrum is too noisy at the corresponding wavelength to allow any useful constraint on the column density.

\subsection{SDSS~J1148$+$5251, $z=6.2575$}

\begin{figure}
   \epsscale{0.93} 
   \centering 
   \plotone{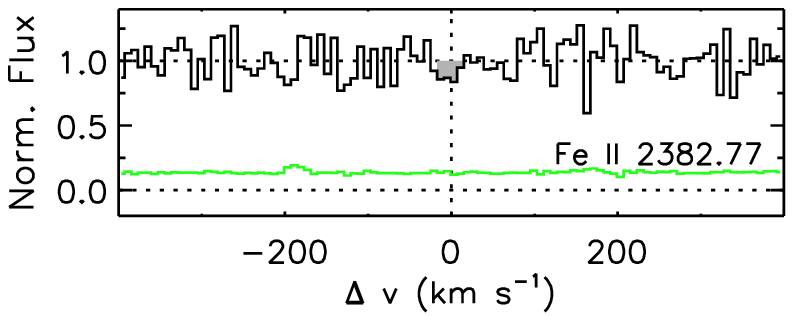}
   \caption{NIRSPEC data covering Fe~II for the $z=6.2575$ system towards SDSS~J1148$+$5251.  The absorption for Fe~II~2383 is not a significant detection.  Lines are as in Figure~\ref{fig:0040_z4.7393}.  Data covering other ions for this system were presented in \citet{becker2011b}. 
     \label{fig:1148_z6.2575}}
\end{figure}

The HIRES data for this system were presented in \citet{becker2006} and re-analyzed in \citet{becker2011b}.  We cover \feii~$\lambda$2382 in the NIRSPEC data, but no significant absorption is detected.  The corresponding region of the spectrum is shown in Figure~\ref{fig:1148_z6.2575}.  

\section{Relative Abundances}\label{sec:abundances}

\begin{deluxetable*}{lccccccccc}
   \tablewidth{\textwidth}
   \centering
   \tablecolumns{10}
   \tablecaption{Relative Abundances of Individual $z > 4.7$ Systems} 
   \tablehead{\colhead{QSO} & 
              \colhead{$z_{\rm abs}$} &
              \colhead{[Si/O]} &
              \colhead{[C/O]} &
              \colhead{[C/Si]} &
              \colhead{[C/Fe]} &
              \colhead{[O/Fe]} &
              \colhead{[Si/Fe]} &
              \colhead{[Mg/Fe]} }
   \startdata
   SDSS~J0040$-$0915  &  4.7393  &  $< 0.33$           &  \nodata           &  $>-0.46$           &  $>-0.12$           &  $>0.02$           &  $ 0.35 \pm 0.06$  &  $>-0.45$           \\
   SDSS~J1208$+$0010  &  5.0817  &  $< 0.27$           &  \nodata           &  $>-0.33$           &  $> 0.14$           &  $>0.20$           &  $ 0.47 \pm 0.07$  &  $  0.28 \pm 0.13$  \\
   SDSS~J0231$-$0728  &  5.3380  &  $ -0.14 \pm 0.06$  &  $-0.42 \pm 0.07$  &  $ -0.29 \pm 0.06$  &  $~~0.13 \pm 0.06$  &  $ 0.55 \pm 0.06$  &  $ 0.42 \pm 0.05$  &  $  0.20 \pm 0.04$  \\
   SDSS~J0818$+$1722  &  5.7911  &  $~~0.00 \pm 0.05$  &  $-0.15 \pm 0.04$  &  $ -0.15 \pm 0.05$  &  $~~0.30 \pm 0.07$  &  $ 0.45 \pm 0.07$  &  $ 0.45 \pm 0.08$  &  \nodata            \\
   SDSS~J0818$+$1722  &  5.8765  &  $ -0.08 \pm 0.07$  &  $-0.17 \pm 0.09$  &  $ -0.09 \pm 0.09$  &  $>-0.02$           &  $>0.15$           &  $>0.07$           &  \nodata            \\
   SDSS~J1148$+$5251  &  6.0115  &  $~~0.04 \pm 0.04$  &  $-0.25 \pm 0.06$  &  $ -0.29 \pm 0.06$  &  $ -0.00 \pm 0.25$  &  $ 0.25 \pm 0.25$  &  $ 0.29 \pm 0.25$  &  \nodata            \\
   SDSS~J1148$+$5251  &  6.1312  &  $ -0.32 \pm 0.24$  &  $-0.65 \pm 0.27$  &  $ -0.34 \pm 0.16$  &  $ -0.09 \pm 0.23$  &  $ 0.56 \pm 0.29$  &  $ 0.25 \pm 0.19$  &  \nodata            \\
   SDSS~J1148$+$5251  &  6.1988  &  $ -0.20 \pm 0.14$  &  $-0.34 \pm 0.17$  &  $ -0.14 \pm 0.12$  &  \nodata            &  \nodata           &  \nodata           &  \nodata            \\
   SDSS~J1148$+$5251  &  6.2575  &  $ -0.02 \pm 0.17$  &  $-0.08 \pm 0.26$  &  $ -0.06 \pm 0.23$  &  $> 0.13$           &  $>0.21$           &  $>0.18$           &  \nodata \\
   \vspace{-0.1in}
   \enddata
   \label{tab:indiv_abundances}
   \tablecomments{Relative abundances are calculated as $[{\rm X/Y}] =
      \log{(N_{\rm X}/N_{\rm Y})} - \log{(n_{\rm X}/n_{\rm
      Y})_{\odot}}$, without any corrections for dust or ionization
      effects.  The column densities used are those integrated over
      the velocity interval spanned by O~I (see
      Table~\ref{tab:columns}).  Solar values are from
      \citet{asplund2009}.}
\end{deluxetable*}

The \lya\ forest at $z \gtrsim 5$ is too highly absorbed to allow accurate \hi\ column density measurements.  At $z \gtrsim 6$, the forest becomes almost completely opaque, and \hi\ cannot be measured at all.  We are therefore unable to directly estimate metallicities for our systems.  We can, however, estimate element ratios based solely on the ionic column densities.  

Directly converting from an ionic column density to a total element column density ($\log{N_{\msiii}}$ to $\log{N_{\rm Si}}$, for example) requires first that there is minimal depletion onto dust grains, and second, that the ionization corrections are known.  Dust-depletion in DLAs tends to increase with metallicity, but is negligible at $[{\rm Fe/H}] \lesssim -2$ \citep{vladilo2004}.  The trend of declining DLA metallicity with redshift \citep[e.g.,][]{wolfe2005} suggests that our $z > 4.7$ systems may be in this very metal poor regime.  Nevertheless, differential dust depletion is a potential source of error in determining relative abundances, an issue to which we will return below.

Ionization corrections are a potentially greater cause for concern.  In the simplest case, the corrections will be small and can be safely neglected.  This is true for heavily self-shielded systems, such as DLAs,  in which nearly all of the hydrogen is in \hi, and nearly all of the metals are in their neutral or first excited states.  Oxygen has a first ionization potential very similar to that of hydrogen, and would therefore expected to be present as \oi.  Silicon, carbon, iron, and magnesium, meanwhile, have first ionization potentials below that of hydrogen, and will not be shielded by neutral gas.  These should therefore be singly ionized \citep[for discussions of ionization corrections in DLAs see, e.g.,][]{prochaska2002b,pettini2008}.   For sub-DLAs ($19.0 \le \log{N_{\mhi}} < 20.3$), the \hi\ self-shielding is significantly reduced, and ionization corrections up to $\sim$0.3 dex for an individual ion have been estimated \citep[e.g.,][]{peroux2007}.  An important question, therefore, is whether significant ionization effects may be present in our $z > 4.7$ systems.  We unfortunately do not have useful constraints on any intermediate ionization states (e.g., \ion{Si}{3}, \ion{Fe}{3}, or \ion{Al}{3}) because these transitions are either lost in the forest, are not covered by our data, or are too weak to be detected.  We note, however, that when inferring relative abundances, it is the difference in the ionization correction between two elements that matters.  Although different elements will be ionized at different rates, the effects will almost always act in the same sense, which is to reduce the population of the ``low'' (\oi, \siii, \cii, etc.) ionization state.  Even for a partially ionized system, therefore, the ratio of two ions (\siii\ and \feii, for example) may be expected to be reasonably close to the total element ratio.

For this work, we will use the fact that all means of affecting ionic column density ratios--dust depletion, ionization effects, as well as intrinsic variations in the relative abundances--will then tend to increase the scatter between systems.  The presence of \oi, which has an ionization potential close to 1~Ryd, already signals that ionization corrections for our $z > 4.7$ systems may be small.  We will argue, moreover, that the low level of scatter in the column density ratios, and their agreements with the ratios in lower-redshift, metal-poor DLAs, strongly suggests that they reflect the underlying relative abundances.

\begin{figure*}
   \epsscale{1.0} 
   \centering 
   \plotone{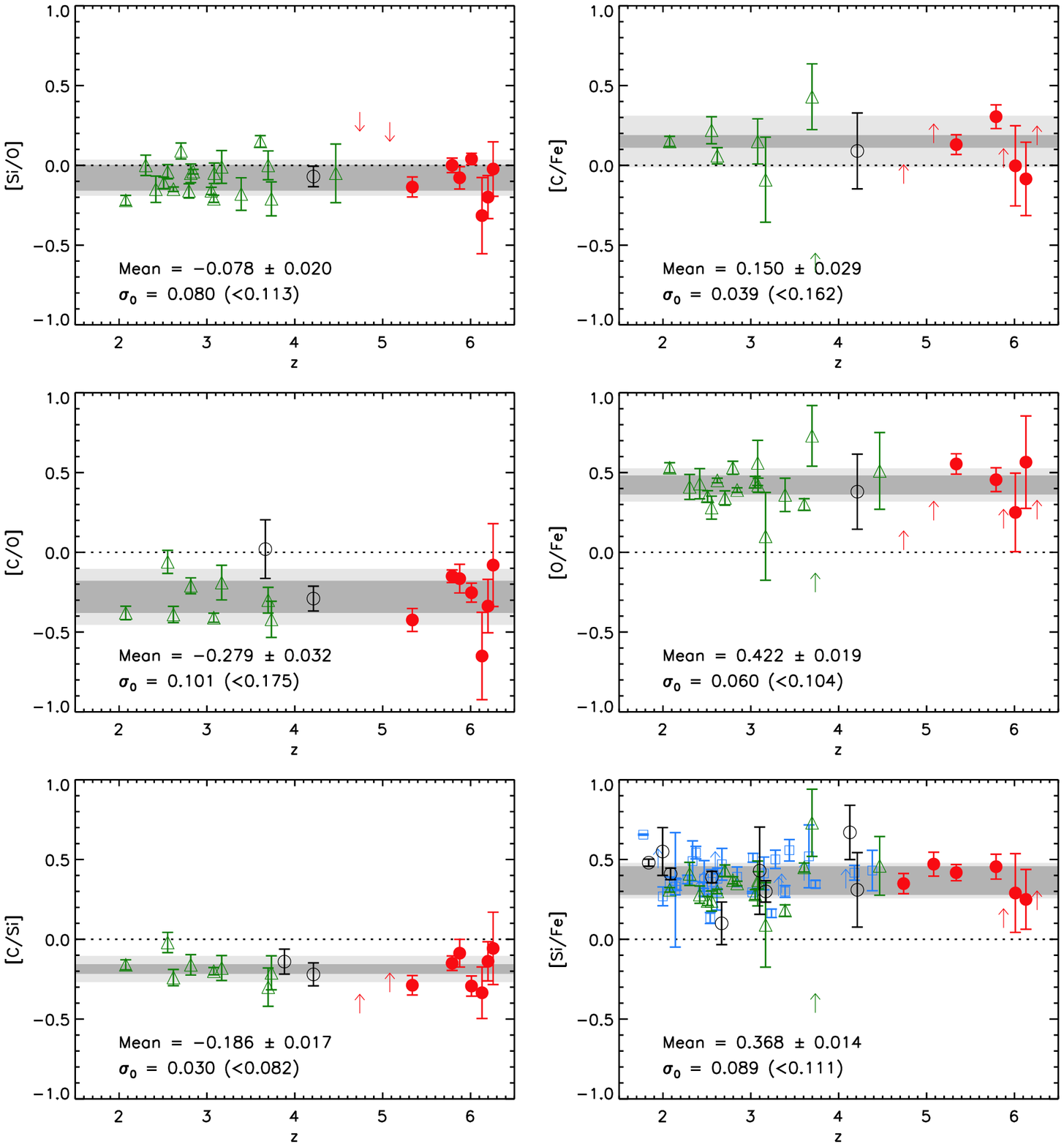}
   \caption{Uncorrected relative abundances in low-ionization systems as a function of redshift.  In all cases, the relative abundances are determined directly from the column density measurements, and are not corrected for dust depletion or ionization effects.  Filled circles are from this work.  Open circles are sub-DLAs from \citet{d-z2003} and \citet{peroux2007}.  Triangles are metal-poor DLAs from \citet{cooke2011b}, and references therein.  Squares are DLAs with $[{\rm M/H}] \le -1.0$ from \citet{wolfe2005}.  Arrows are 1-$\sigma$ upper or lower limits, which at $z \ge 4.7$ are for this work and otherwise are for literature values.  The mean value with a 1-$\sigma$ uncertainty is given in each panel, where the values have been calculated from all measurements (see Table~\ref{tab:mean_abundances}).  Estimates of the nominal intrinsic r.m.s. scatter, $\sigma_{0}$, and the 95\% upper limit on $\sigma_{0}$ (in parentheses) are shown as dark and light shaded bands, respectively.
        \label{fig:abundances}}
\end{figure*}

The uncorrected relative abundances  for individual $z > 4.7$ systems are given in Table~\ref{tab:indiv_abundances}.  These do not include any corrections for dust depletion or ionization effects.  We compare the $z > 4.7$ results to values for lower-redshift systems in Figure~\ref{fig:abundances}.  The lower-redshift data include three groups of low-ionization absorption systems measured from high-resolution data, for which accurate column density measurements for some of the most abundant ions (\oi, \siii, \cii, and \feii) are available.  The first group is a compilation of very metal-poor DLAs from \citet{cooke2011b}, which includes their own measurements and other results from the literature (see references therein).  These DLAs were selected to have very low metallicities ($[{\rm Fe/H}] \le -2.0$), and hence weak absorption lines.  In particular, column densities have been measured for \oi\ and \cii, whose transitions are normally saturated for higher metallicity DLAs \citep[e.g.,][]{prochaska2002a}.  We omit the potentially carbon-enhanced metal-poor DLA from \citet{cooke2011a} due to possible large uncertainties in the column densities.  The velocity width of this system is sufficiently narrow ($b = 2.36$~\kms) that thermal broadening may be significant.  The fit is largely driven by the multiple lines of \siii, for which the corresponding temperature would be $\sim$9500~K.  In a thermally broadened model, the \cii\ column density would be substantially ($\gtrsim 1$ dex) lower than in the turbulent model they assume (confirmed by R. Cooke, private communication).  We note that thermal broadening is much less likely to be an issue for our systems, as our narrowest absorber has $b = 3.8$~\kms, for which the corresponding temperature for \siii\ would be $\sim$25000~K.  We further compare our $z > 4.7$ data to a sample of DLAs from \citet{wolfe2005} with metallicities $-2.7 \le [{\rm Si/H}] \le -1.0$.  These have column densities measured for \siii\ and \feii, which have weak transitions that remain optically thin.  The upper cutoff in metallicity is chosen to avoid systems with an enhancement in [Si/Fe] that is most plausibly due to dust depletion \citep{wolfe2005}.  Finally, we include a selection of sub-DLAs from \citet{d-z2003} and \citet{peroux2007}.  For these systems we restrict ourselves to measurements where the metal lines fall redward of the \lya\ forest, are not strongly saturated, and are free from any apparent contamination from other lines.  We have not included the metal-poor DLA measurements from \citet{penprase2010}, since these were made from medium-resolution spectra.  Although they are broadly consistent with the other literature values, the trends in relative abundances with metallicity identified by Penprase et al. could constitute an intriguing counterexample to the results described below if confirmed with high-resolution data.

\begin{deluxetable}{lccc}
   \tablewidth{0.4\textwidth}
   \centering
   \tablecolumns{4}
   \tablecaption{Mean Relative Abundances} 
   \tablehead{\colhead{Group} &
              \colhead{\# Systems} &
              \colhead{Mean} &
              \colhead{$\sigma_{0}$\tablenotemark{a}} }
   \startdata
   \cutinhead{[Si/O]}
   sub-DLAs	      &   1  &  \nodata           &  \nodata            \\
   VMP-DLAs           &  19  &  $-0.08 \pm 0.02$  &  0.09 ($\le 0.13$)  \\
   $z > 4.7$ Systems  &   7  &  $-0.04 \pm 0.03$  &  0.05 ($\le 0.18$)  \\
   All                &  27  &  $-0.08 \pm 0.02$  &  0.08 ($\le 0.11$)  \\ [-1ex]
   \cutinhead{[C/O]}
   sub-DLAs	      &   2  &  \nodata           &  \nodata            \\
   VMP-DLAs           &   8  &  $-0.30 \pm 0.05$  &  0.11 ($\le 0.22$)  \\
   $z > 4.7$ Systems  &   7  &  $-0.26 \pm 0.05$  &  0.09 ($\le 0.30$)  \\
   All                &  17  &  $-0.28 \pm 0.03$  &  0.10 ($\le 0.18$)  \\ [-1ex]
   \cutinhead{[C/Si]}
   sub-DLAs           &   2  &  \nodata           &  \nodata            \\
   VMP-DLAs           &   8  &  $-0.18 \pm 0.02$  &  0.04 ($\le 0.12$)  \\
   $z > 4.7$ Systems  &   7  &  $-0.21 \pm 0.04$  &  0.04 ($\le 0.17$)  \\
   All                &  17  &  $-0.19 \pm 0.02$  &  0.03 ($\le 0.08$)  \\ [-1ex]
   \cutinhead{[C/Fe]}
   sub-DLAs           &   1  &  \nodata           &  \nodata            \\
   VMP-DLAs           &   6  &  $ 0.14 \pm 0.03$  &  0.04 ($\le 0.29$)  \\
   $z > 4.7$ Systems  &   4  &  $ 0.17 \pm 0.07$  &  0.09 ($\le 0.47$)  \\
   All                &  11  &  $ 0.15 \pm 0.03$  &  0.04 ($\le 0.16$)  \\ [-1ex] 
   \cutinhead{[O/Fe]}
   sub-DLAs           &   1  &  \nodata           &  \nodata            \\
   VMP-DLAs           &  12  &  $ 0.39 \pm 0.02$  &  0.04 ($\le 0.08$)  \\
   $z > 4.7$ Systems  &   4  &  $ 0.50 \pm 0.05$  &  0.00 ($\le 0.36$)  \\
   All                &  17  &  $ 0.40 \pm 0.02$  &  0.04 ($\le 0.08$)  \\ [-1ex]
   \cutinhead{[Si/Fe]}
   sub-DLAs           &   9  &  $ 0.40 \pm 0.04$  &  0.09 ($\le 0.24$)  \\
   Metal-poor DLAs    &  31  &  $ 0.38 \pm 0.02$  &  0.10 ($\le 0.13$)  \\
   VMP-DLAs           &  17  &  $ 0.33 \pm 0.02$  &  0.06 ($\le 0.10$)  \\
   $z > 4.7$ Systems  &   6  &  $ 0.41 \pm 0.03$  &  0.00 ($\le 0.12$)  \\
   All                &  63  &  $ 0.37 \pm 0.01$  &  0.09 ($\le 0.11$)  \\
   \vspace{-0.1in}
   \enddata
   \label{tab:mean_abundances}
   \tablecomments{Relative abundances are computed assuming no dust
   depletion or ionization corrections.  Mean values are given only
   for samples with $\ge 4$ systems.}
   \tablenotetext{a}{Estimated intrinsic scatter in the relative
   abundances.  The 95\% confidence upper limit on $\sigma_0$ is given
   in parentheses.}
\end{deluxetable}

The mean uncorrected relative abundances for the $z > 4.7$ and lower-redshift systems are given in Table~\ref{tab:mean_abundances}.  We give the relative values for all six combinations of our four measured elements, though these are obviously not all independent.   Limits are not included, though these are generally consistent with the mean values.  We also estimate the intrinsic scatter in the relative abundances by assuming a log-normal distribution and computing the r.m.s. logarithmic scatter, $\sigma_0$, which, when added in quadrature to the measurement errors, produces a reduced $\chi^2$ equal to 1.0 for a fit to a constant value.  This nominal value of $\sigma_0$ is added to the individual errors when computing the mean values.  A 95\% upper limit on $\sigma_0$ is also estimated as that which produces the 95\% lower bound on $\chi^2$ for the available number of measurements.  

Two points stand out from Figure~\ref{fig:abundances} and Table~\ref{tab:mean_abundances}.  First, the relative abundances are very similar among different groups of absorbers.  For a given element ratio, all of the mean values agree to within $\lesssim 2\sigma$, and most agree to within $\lesssim 1\sigma$.  Second, the intrinsic scatter for all element ratios appears to be small.  The nominal intrinsic scatter within the complete sample is always $\le 0.1$~dex (r.m.s.).  Even the 95\% upper limits on $\sigma_0$ are generally $\lesssim 0.1$.  The exceptions are [C/O], for which both \cii\ and \oi\ are measured from only one or two strong lines, and [C/Fe], where the number of data points with small errors is low.  Values of $\sigma_0$ larger than 0.1 are also generally allowed for individual groups of systems only when there are relatively few measurements and/or the measurement error bars are large, and so adding intrinsic scatter has less impact on the total scatter.   

We emphasize that the intrinsic scatter will include any variations in dust depletion and ionization corrections, and should therefore give an upper limit on the true variation in the element ratios.  The intrinsic scatter may also be overestimated if the measurement uncertainties are too small.  Absorption line measurements are  susceptible to errors caused by contamination from unrelated lines, which may be difficult to detect.  This is a particular problem when the measurements rely on only one or two lines, as is always the case for \oi\ and \cii, and may also be true for \siii\ and \feii, depending on the strength of the system and the wavelength coverage of the data.  Hidden saturation effects, even in high-resolution data, are another potential source of uncertainty that may not be properly included in the formal measurement errors.  These factors all suggest that the minimal scatter we measure in the column density ratios strongly indicates that the variation in the underlying element ratios is genuinely small.  The lack of strong variations, in turn, suggests that the column density ratios are a good indication of the true relative element abundances.

\section{Discussion}\label{sec:discussion}

We have shown that the ionic column density ratios in our $z > 4.7$ systems are remarkably similar to those in low-ionization systems at $z \sim 2$-4.  By necessity, most of the lower-redshift data are drawn from a sample of low-metallicity DLAs; these have relatively weak metal lines, making it possible to measure  column densities for \oi\ and \cii.  While the metal-poor systems are a convenient reference sample, it should be remembered that these were deliberately drawn from the tail of the metallicity distribution, and may not represent typical DLAs at these redshifts (although typical DLAs with metallicities up to $[{\rm Si/H}] = -1.0$ are included in the [Si/Fe] measurements).  In contrast, our $z > 4.7$ sample includes nearly all of the low-ionization systems detected in an unbiased survey \citep[][and additional observations of $z \ge 5$ quasars by our group]{becker2006,becker2011b}.  We turn now to consider the possible origins of these high-redshift metals, and to compare our results with the relative abundances in metal-poor stars.

\subsection{Pop II or Pop III Enrichment?}

It is worth considering whether our high-redshift systems reflect the yields of conventional Pop II stars, or perhaps an early generation of metal-free stars.  \citet{cooke2011b} noted that the relative abundances in their metal-poor DLAs were consistent with enrichment from either low-metallicity ($Z  \le 1/3\; Z_{\odot}$) Pop II stars or zero-metallicity ``Pop III'' stars (with masses of 10 to 100~M$_\odot$), based on comparisons with theoretical yields.  This suggests that our systems could also be dominated by metals from either population.  There are several issues to note, however.  First, much of the discriminating power in the lower-redshift systems came from the abundances of nitrogen and aluminum, which we are unfortunately unable to measure.\footnote{The transitions for N~I and N~II fall in the \lya\ forest, which is highly absorbed at $z > 4.7$.  While Al~II~$\lambda$1670 can in principle be measured for our systems, in nearly all cases where we had spectral coverage the data were of too low quality to place meaningful limits.}  Equally important, however, are the uncertainties in the theoretical yields.  These depend on a broad array of minimally constrained factors, including mixing, mass loss, and the mechanics of supernovae.  The yields are also highly sensitive to stellar mass \citep[e.g.,][]{woosley1995,heger2010}, which means that interpolating over a sparsely sampled mass grid \citep[e.g.,][]{chieffi2004,nomoto2006} may not give accurate mean abundances.  In addition, the  results will depend on the lower and upper mass limits over which the yields are integrated.

We suggest that the abundance ratios in our high-redshift systems may be most useful as empirical constraints on the integrated yields of (presumably low metallicity) massive Pop II stars.  Although the first generation of stars must necessarily be metal-free, the phase of cosmic star formation dominated by Pop III stars is expected to be brief, as high-density regions will quickly become enriched above the critical metallicity threshold.  The transition to Pop II stars should occur well before $z \sim 10$ \citep[e.g.,][]{maio2010}.  Current estimates of the buildup of stellar mass at $z > 4$ suggest that only $\sim$1/3 of the stellar mass at $z \sim 6$ would have been in place by $z \sim 10$ \citep{gonzalez2011}.  It is therefore reasonable to assume that Pop II stars have dominated the metal enrichment by $z \sim 6$.  As noted above, our high-redshift systems represent typical metal-enriched regions of the universe at $z \sim 5$-6, at least when weighting by cross section.  Their relative abundances are therefore likely to reflect the integrated yields of Pop II stars, and massive stars in particular, since there has not been enough time for low-mass stars to evolve.  Over longer timescales, contributions from lower-mass stars through delayed Type Ia supernova (which will add iron) and AGB winds (which will add carbon) will start to become important.  The similarity in the abundances between our $z > 4.7$ systems and those at $2 < z < 4.5$, however, suggests that supernovae of massive stars remain the dominant source of enrichment even at lower redshifts.  This may either be because the moderate redshift, low-metallicity systems were enriched at earlier times, or because they were enriched relatively recently.

\subsection{Comparison with Metal-Poor Stars}

\begin{figure}
   \epsscale{1.15} 
   \centering 
   \plotone{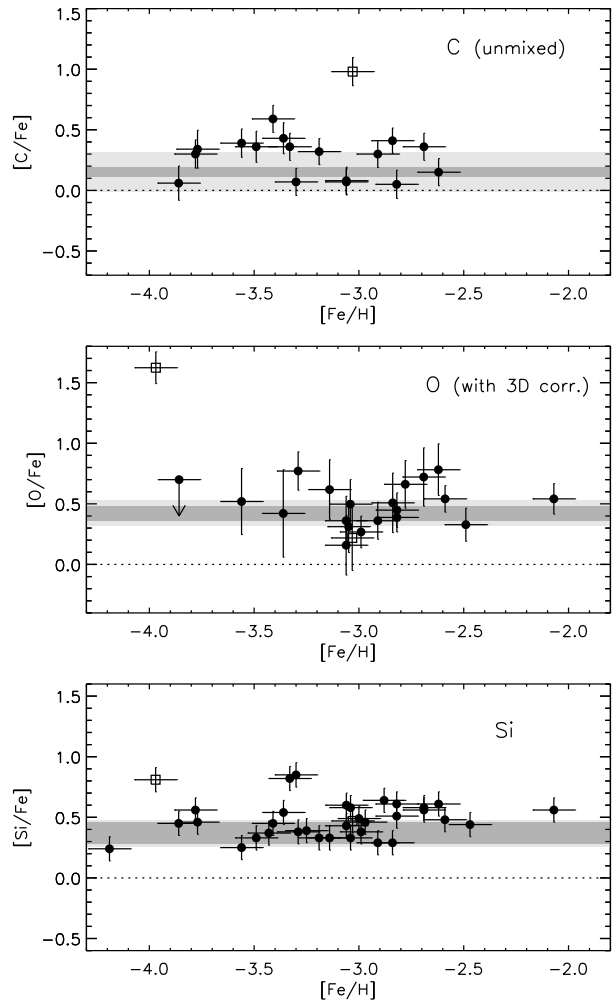}
   \caption{Ratios of carbon, oxygen, and silicon to iron for the sample of metal-poor stars from \citet{cayrel2004}.  All stellar values have been adjusted to \citet{asplund2009} solar values.  For carbon, only ``unmixed'' stars are shown \citep{spite2005}.  Values for oxygen include corrections for three-dimensional effects (see notes in Cayrel et al 2004).  The carbon-enhanced stars CS 22949-037 and CS 22892-052 are plotted with open squares.  In each panel, the values for low-ionization quasar absorption line systems are shown as shaded bands.  Dark grey bands span the nominal 1-$\sigma$ r.m.s. intrinsic scatter ($\sigma_0$) around the mean value in the absorption-line data, while light grey bands reflect the 95\% upper limit on $\sigma_0$ (see Figure~\ref{fig:abundances} and text for details).
     \label{fig:stars}}
\end{figure}

Finally, we can compare the abundances in the high-redshift absorbers to those in metal-poor stars.  Figure~\ref{fig:stars} shows the ratios of C, O, and Si to Fe as a function of metallicity for a sample of very metal-poor halo giant stars from \citet{cayrel2004}.  For each element, we also plot the estimated intrinsic 1-$\sigma$ range in [X/Fe] spanned by the quasar absorption line systems.  The agreement is generally good, which is consistent with the picture in which present-day metal-poor stars were potentially formed at very high redshifts.  It should be noted, however, that the stellar values depend on the choice of model parameters.  The stellar carbon abundances were determined from the G-band of CH at $\sim$4300~\AA.  These are 1D/LTE values, although the (negative) corrections due to 3D and non-LTE effects are potentially large  \citep{asplund2005a,collet2006,frebel2008}.  Atmospheric carbon abundances in giant stars can be altered by convective mixing.  We therefore plot [C/Fe] values only for those stars that appear to be unmixed based on their relative abundances of C, N, and Li \citep{spite2005}.  Oxygen was measured from the forbidden [\oi] 6300~\AA\ line, which is not expected to suffer non-LTE effects.  Three-dimensional effects can be significant at low metallicities, however, and we plot the \citet{cayrel2004} values for which a 1D to 3D correction of $\sim -0.3$ has been applied, following the results of \citet{nissen2002}.  Silicon abundances were computed from a single line at 4103~\AA, and are 1D/LTE.

The C/O ratio is of particular interest as a diagnostic of the IMF, as more massive stars tend to produce more oxygen relative to carbon.  An added advantage is that both carbon and oxygen are formed during the hydrostatic burning phase, which helps to simplify the comparison with theoretical yields.  The most direct constraints on the IMF can potentially be placed at low metallicities, where mass loss from stellar winds is minimal.  The results from stellar abundances, however, again depend somewhat on the choice of model parameters.  \citet{tomkin1992} and \citet{akerman2004} measured [C/O] in metal-poor dwarf and giant stars, respectively, using high-excitation \ci\ and \oi\ lines.  These lines have the advantage that, while the separate C and O measurements both depend strongly on the effective temperature, their ratio generally does not.  Non-LTE effects were included for the dwarf stars by \citet{tomkin1992} and for the giant stars by \citet{fabbian2009b}.  Three-dimensional modeling was not done in either case, but the effects are expected to be small \citep{fabbian2009b}.  Both groups found that $[{\rm C/O}] \sim -0.6$ to $-0.5$ over $-2.0 < [{\rm O/H}] < -0.5$ ($-2.5 < [{\rm Fe/H}] < -1$).  The \citet{akerman2004} sample extends to somewhat lower metallicities, and exhibits an upturn at the low-metallicity end, reaching $[{\rm C/O}] \sim -0.3$ to 0.0 at $[{\rm O/H}] \sim -3$, depending on how the non-LTE corrections are calculated \citep{fabbian2009b}.  This trend is also seen in the metal poor giants studied by \citet{cayrel2004} and \citet{spite2005}, but only when the negative 3D corrections are applied to the oxygen abundances (and not to the carbon).

The apparent upturn in stellar [C/O] values at the lowest metallicities has been cited as evidence of early enrichment from Pop III stars and/or a top-heavy IMF \citep{akerman2004}.  Very metal-poor \citep{cooke2011b} and very high-redshift (this work) quasar absorption line systems have $[{\rm C/O}] \sim -0.3$, similar to the most metal-poor stars.  The implications for the IMF, however, will depend on the adopted theoretical yields \citep[see also the discussion in][]{cooke2011b}.  For example, $[{\rm C/O}] \sim -0.3$ is close to the value calculated by \citet{cescutti2009} for a standard \citet{kroupa2002} IMF corrected for binaries, using the yields of \citet{meynet2002}, which include the effects of rotation and mass loss \citep[these yields were also used by][]{akerman2004}.  Based on this calculation, the [C/O] ratios seen in quasar absorption line systems and the most metal-poor stars show no evidence for a top-heavy IMF.  In that case, however, it becomes challenging to explain why even lower [C/O] ratios are seen in galactic halo and bulge stars with $-2.0 < [{\rm O/H}] < -0.5$.  Perhaps the most important point is that the mean [C/O] ratio in the quasar absorption line systems is significantly higher than the minimum value seen in metal-poor stars.  Unless systematic errors remain in some or all of the measurements, this difference may empirically suggest variations in the IMF, regardless of theoretical yields.

Finally, we emphasize that we have compared our results mainly to metal-poor stars that are not carbon enhanced.  These are underrepresented in the \citet{cayrel2004} sample, yet their generally poor agreement with quasar absorption line abundances is already apparent from the two such stars included in Figure~\ref{fig:stars}.  If carbon-enhanced stars fairly reflect their native ISM abundances, then these abundances are no longer common by $z \sim 6$.  This raises the intriguing possibility that most carbon-enhanced stars were formed at even earlier times.  On the other hand, our high-redshift sample is relatively small, and we could have easily missed a carbon-enhanced population even if it made up $\sim$20\% of the total number density, similar to the frequency among very metal-poor stars \citep{lucatello2006}.  Further surveys for high-redshift and metal-poor absorption line systems should better establish what fraction of absorbers are carbon-enhanced \citep[e.g.,][]{cooke2011a}, and clarify their connection to metal-poor stars.

\section{Conclusions}\label{sec:conclusions}

We have measured the relative abundances of C, O, Si, and Fe in nine low-ionization quasar absorption line systems at $4.7 < z < 6.3$.  The new data presented here are derived from optical and infrared spectra taken with Keck, Magellan, and the VLT, and supplement previous measurements made with high-resolution optical spectra.  The absorption lines are generally weak, which allows us to measure column densities even for carbon and oxygen, which are generally saturated in typical lower-redshift systems.

The relative abundances in our $z > 4.7$ systems agree well with those measured from sub-DLAs and metal-poor DLAs at lower redshifts ($2 \lesssim z \lesssim 4.5$).  The lower-redshift sample includes very metal poor systems, where carbon and oxygen abundances can be measured \citep[][and references therein]{cooke2011b}.  Among our systems and the literature results there is remarkably little apparent intrinsic scatter ($\sigma_0 \lesssim 0.1$~dex) in the column density ratio of any two of the dominant ions (\cii, \oi, \siii, and \feii), and no apparent evolution with redshift.  The lack of scatter suggests that the relative element abundances are correctly estimated from the ratios of the corresponding ions, with minimal ionization effects or dust depletion.  It also suggests that the metal inventories of most metal poor absorption systems over $2 < z < 6$ are dominated by a similar stellar population.  At $z \sim 6$, the finite age of the universe means the metals must have been produced by short-lived, massive stars.  A reasonable interpretation, therefore, is that the relative abundances in metal-poor quasar absorption-line systems reflect the integrated yields from prompt supernovae (Type II, Ib/c, and prompt Ia), with minimal contribution from delayed Type Ia supernovae or AGB winds.  From these data we can directly infer, for example, that massive stars yield an iron abundance with respect to oxygen that is roughly one third of the solar value.

The measurements presented here should help in understanding the nature of stellar populations at both high and low redshifts.  The lack of exotic abundance patterns at $z \sim 5$-6 suggests that ordinary Pop II (rather than massive Pop III) stars are likely to have produced most of the ionizing photons during hydrogen reionization.  The broad agreement between the abundances in our high-redshift absorption systems and those in metal-poor halo stars \citep[e.g.,][]{cayrel2004} further provides a direct link between these stars and the high-redshift material out of which they are likely to have formed.  Future measurements at high redshifts should help to strengthen this connection between the oldest stars in our galaxy and star formation in the early universe, as well as reveal the metal abundances present at the earliest stages of galactic chemical evolution.

\acknowledgments

We thank Ryan Cooke, John Eldridge, John Norris, Max Pettini, and Andrew McWilliam for many helpful discussions throughout the course of this work.  We also thank the referee for their helpful comments.  We wish to recognize and acknowledge the very significant cultural role and reverence that the summit of Mauna Kea has always had within the indigenous Hawaiian community.  We are most fortunate to have the opportunity to conduct observations from this mountain.  GB has been supported by the Kavli Foundation.  WS received support from the National Science Foundation through grant AST 06-06868.  MR received support from the National Science Foundation through grant AST 05-06845.

\bibliographystyle{apj}
\bibliography{/Users/gdb/tex/refs}

\clearpage
\begin{landscape}
\begin{deluxetable*}{lcccccccccc}
   \tablewidth{\textwidth}
   \centering
   \tablecolumns{11}
   \tablecaption{Column Densities} 
   \tablehead{\colhead{QSO} & 
              \colhead{$z_{\rm abs}$} &
              \colhead{$\log{N_{\mhi}}$} &
              \colhead{$\log{N_{\mcii}}$} &
              \colhead{$\log{N_{\moi}}$} &
              \colhead{$\log{N_{\mmgii}}$} &
              \colhead{$\log{N_{\msiii}}$} &
              \colhead{$\log{N_{\mfeii}}$} &
              \colhead{$v_{\rm lo},v_{\rm hi}$\tablenotemark{a}} &
              \colhead{$b$\tablenotemark{b}} &
              \colhead{Ref.\tablenotemark{c}} \\
               & & & & & & & & (${\rm km\,s^{-1}}$) & (${\rm km\,s^{-1}}$) & }
   \startdata
   SDSS~J0040$-$0915  &  4.7393                   &  $<20.4$       &  $>14.6$           &  $>15.0$           &  $>13.4$           &  $14.13 \pm 0.02$                   &  $13.77 \pm 0.06$  &  -34,80               &  \nodata        &  1    \\ 
   SDSS~J1208$+$0010  &  5.0817                   &  $<20.3$       &  $>14.3$           &  $>14.7$           &  $13.65 \pm 0.11$  &  $13.75 \pm 0.03$                   &  $13.27 \pm 0.07$  &  \nodata              &  $9.5 \pm 0.2$  &  1    \\
   SDSS~J0231$-$0728  &  ~5.3380\tablenotemark{d}  &  $<20.1$       &  $13.79 \pm 0.05$  &  $14.47 \pm 0.05$  &  $13.02 \pm 0.01$  &  $13.15 \pm 0.04$                   &  $12.73 \pm 0.04$  &  -25,36 (-38,49)      &  \nodata        &  1    \\
                      &                           &                &  $14.13 \pm 0.04$  &                    &  $13.38 \pm 0.01$  &  $13.42 \pm 0.02$                   &  $12.94 \pm 0.05$  &  -119,160 (-147,184)  &  \nodata        &       \\
   SDSS~J0818$+$1722  &  ~5.7911\tablenotemark{d}  &  \nodata       &  $14.13 \pm 0.03$  &  $14.54 \pm 0.03$  &  \nodata           &  $13.36 \pm 0.04$                   &  $12.89 \pm 0.07$  &  -86,45 (-100,60)     &  \nodata        &  1,2  \\
                      &                           &                &  $14.19 \pm 0.03$  &                    &  \nodata           &  $13.41 \pm 0.03$                   &  $12.95 \pm 0.07$  &  -124,45 (-140,60)    &  \nodata        &       \\
   SDSS~J0818$+$1722  &  5.8765                   &  \nodata       &  $13.62 \pm 0.07$  &  $14.04 \pm 0.05$  &  \nodata           &  $12.78 \pm 0.05$                   &  $<12.7$           &  -16,12 (-30,25)      &  \nodata        &  1,2  \\
   SDSS~J1148$+$5251  &  6.0115                   &  \nodata       &  $14.14 \pm 0.06$  &  $14.65 \pm 0.02$  &  \nodata           &  $13.51 \pm 0.03$                   &  $13.21 \pm 0.25$  &  -36,32 (-50,45)      &  \nodata        &  1,2  \\
   SDSS~J1148$+$5251  &  6.1312                   &  \nodata       &  $13.88 \pm 0.15$  &  $14.79 \pm 0.23$  &  \nodata           &  $13.29 \pm 0.06$                   &  $13.03 \pm 0.18$  &  \nodata              &  $5.7 \pm 0.7$  &  1,2  \\
   SDSS~J1148$+$5251  &  6.1988                   &  \nodata       &  $12.90 \pm 0.11$  &  $13.49 \pm 0.13$  &  \nodata           &  $12.12 \pm 0.05$                   &  \nodata           &  -12,10               &  \nodata        &  1,2  \\
   SDSS~J1148$+$5251  &  6.2575                   &  \nodata       &  $13.78 \pm 0.21$  &  $14.12 \pm 0.15$  &  \nodata           &  $~12.92 \pm 0.08$\tablenotemark{e}  &  $<12.7$           &  \nodata              &  $3.8 \pm 0.3$  &  1,2  \\
   \vspace{-0.1in}
   \enddata
   \label{tab:columns}
   \tablenotetext{a}{Minimum and maximum velocities, with respect to
   the nominal redshift, over which the optical depths were integrated
   for systems measured using the Apparent Optical Depth method.
   Where two sets of numbers are given, the first set refers to the
   HIRES data, and the set in parentheses refers to the
   lower-resolution (X-Shooter or NIRSPEC) data.}
   \tablenotetext{b}{Doppler $b$-parameter for systems fit using a
   Voigt profile}
   \tablenotetext{c}{References for column density measurements:
   1--This work, 2--\citet{becker2011b}}
   \tablenotetext{d}{The first line gives the column densities
   integrated over the velocity interval spanned by O~I, while the
   second line gives the total column densities integrated over the
   entire interval where Si~II and C~II are detected.}
   \tablenotetext{e}{See notes in text.}
\end{deluxetable*}
\clearpage
\end{landscape}

\end{document}